\documentclass[preprintnumbers,amsmath,amssymb]{revtex4}
\usepackage{epsfig}
\usepackage{color}
\usepackage{dsfont}
\usepackage[colorlinks,urlcolor=blue,linkcolor=blue,citecolor=red]{hyperref}
\begin{document}
\setlength{\voffset}{1.0cm}
\title{Peierls instability for systems with several Fermi surfaces: \\ an example from the chiral Gross-Neveu model}
\author{Michael Thies\footnote{michael.thies@gravity.fau.de}}
\affiliation{Institut f\"ur  Theoretische Physik, Universit\"at Erlangen-N\"urnberg, D-91058, Erlangen, Germany}
\date{\today}

\begin{abstract}
As is well known, the chiral Gross-Neveu model at finite density can be solved semi-classically with the help of the chiral spiral mean field.
The fermion spectrum has a single gap right at the Fermi energy, a reflection of the Peierls instability.
Here, we divide the $N$ fermion flavors up into two subsets to which we attribute two different densities. The Hartree-Fock
ground state of such a system can again be found analytically, using as mean field the ``twisted kink crystal" of Basar and
Dunne. Its spectrum displays two gaps with lower edges coinciding with  the two Fermi energies. This solution is favored over the
homogeneous one, providing us with an explicit example of a multiple Peierls instability.
\end{abstract}

\maketitle

\section{Introduction} 
\label{sec1}

Peierls distortion in one-dimensional condensed matter systems refers to a modulation of the distance between 
neighboring ions and its effect on the electron structure \cite{L1}. In this form, there seems to be no obvious analogue in particle physics or relativistic quantum field theory.
However, the physics reason behind the Peierls distortion is the opening of a gap at the Fermi surface of the electrons. 
As far as this aspect of the ``Peierls instability" is concerned, it is also ubiquitous in one-dimensional relativistic field theory models. The simplest example is perhaps the mass 
gap dynamically generated in theories of self-interacting, massless fermions. This is nothing but a gap at the top of the filled Dirac sea.
At finite density, the Fermi energy and thus the gap are expected to move upwards in energy. This is achieved in many cases by generating an inhomogeneous, periodic mean field
with a gapped spectrum  -- the fermions form a crystal. Numerous examples have been found, notably also in the context of thermodynamics at finite temperature and chemical potential. 

A family of toy models where such questions have been addressed and often answered analytically are Gross-Neveu (GN) type models, Dirac fermions in 1+1 dimensions with quartic
self-interactions in the limit of a large number of flavors.
The two most important representatives are the original GN model with discrete Z$_2$ chiral symmetry and Lagrangian \cite{L2}
\begin{equation}
{\cal L} = \bar{\psi} i \partial \!\!\!/ \psi + \frac{g^2}{2} (\bar{\psi}\psi)^2 
\label{1.1}
\end{equation}
and the chiral version thereof with continuous U(1) chiral symmetry and Lagrangian
\begin{equation}
{\cal L} = \bar{\psi} i \partial \!\!\!/  \psi + \frac{g^2}{2} \left[(\bar{\psi}\psi)^2+ (\bar{\psi}i \gamma_5 \psi)^2 \right],
\label{1.2}
\end{equation}
also referred to as two-dimensional Nambu--Jona-Lasinio (NJL$_2$) model \cite{L3}. In Eqs.~(\ref{1.1},\ref{1.2}), flavor labels have been suppressed as usual
($\bar{\psi}\psi = \sum_{k=1}^N \left(\bar{\psi}_k \psi_k \right)$ etc.)

In the GN model, matter at finite density illustrates the Peierls instability in a very clear fashion. Whereas the mean field in the vacuum is just a constant mass, it becomes periodic
at any finite density. The self-consistent mean fields which have been found are of Jacobi elliptic type, closely related to the
non-relativistic Lam\'e equation. The same functional form that has been found at zero temperature was shown to be self-consistent also at any finite temperature and chemical potential.
For a summary of these results within the Dirac Hartree-Fock (HF) approximation appropriate for the large $N$ limit, see Ref.~\cite{L4}.

By contrast, in the massless NJL$_2$ model, one gets away with a much simpler functional form of the mean field, the complex chiral spiral \cite{L5}. The HF potential assumes
the form of a plane wave, $m \exp\{2i\mu x\}$. As a consequence, at zero temperature the mass gap is simply moving rigidly up in energy with increasing chemical potential $\mu$,
a manifestation of the U(1) anomaly. Equilibrium thermodynamics can also be handled by allowing the radius $m$ of the chiral spiral to vary with temperature in exactly the same way as the fermion
mass at zero chemical potential \cite{L6}.

Basar and Dunne have discovered and explored yet another elliptic finite gap  potential for which the Dirac-HF equation (or, equivalently,
the Bogoliubov-deGennes (BdG) equation) can be solved in closed analytical form \cite{L7,L8}. It is actually more involved than both the ``cnoidal" potentials of the GN model and the chiral spiral
of the NJL$_2$ model. It is not related to the Lam\'e equation, but based on Weierstrass elliptic functions. Remarkably, it contains all known self-consistent
potentials of the GN models (both real and complex) as special limits of the parameters. An attempt to solve the thermodynamics of the NJL$_2$ model with this potential however
showed that the system prefers parameters leading back to the simpler chiral spiral \cite{L9}, independently of temperature and chemical potential.
Thus, in that case, we have a sophisticated candidate for the NJL$_2$ mean field,
but do not yet understand what it describes physically, at least if one aims at exhausting its full parameter space.

More recently, finite $N$ studies have confirmed some of the semiclassical findings, either by using numerical lattice techniques \cite{L10} (even down to $N=2$ \cite{L11}) or by
exploiting analytically integrability of GN models (relation to Wess-Zumino-Witten model \cite{L12}, Bethe Ansatz \cite{L13}).
A new twist of the recent study \cite{L14} has been to divide the $N$ fermion flavors up into two groups ($N-a$ called ``neutral" and $a$ called ``charged"), introducing a chemical
potential for charged fermions only. The fraction of charged fermions, $a/N$, is a very useful parameter indeed. It gives rise to additional freedom in the standard GN model
without destroying its solvability. A closer look at the semiclassical part of Ref.~\cite{L14} shows that the authors are successful not with the 
original mean field for the massless GN model \cite{L15}, but with the potential used previously to solve the massive model \cite{L16}, although they are dealing with the massless one.
The massive GN model differs from the massless one by a bare Dirac mass term $\sim m_0$,
\begin{equation}
{\cal L} = \bar{\psi} i (\partial \!\!\!/ - m_0)  \psi + \frac{g^2}{2} \left[(\bar{\psi}\psi)^2+ (\bar{\psi}i \gamma_5 \psi)^2 \right].
\label{1.3}
\end{equation}


\begin{figure}
\begin{center}
\epsfig{file=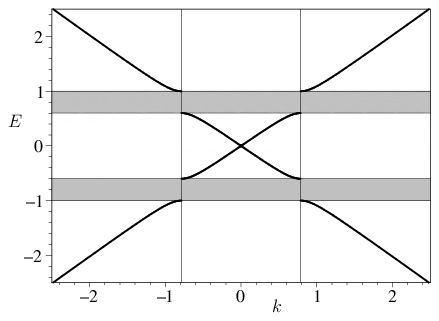,height=5cm,angle=0}\ \ \epsfig{file=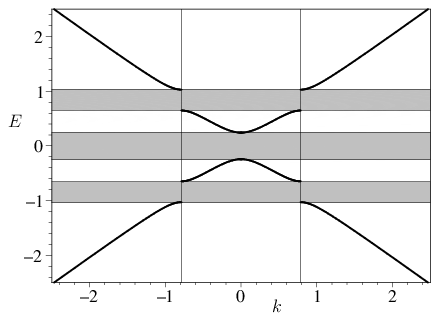,height=5cm,angle=0}
\caption{Examples of  fermion dispersion relation of massless (left) and massive (right) GN models at finite density. Shaded stripes: band gaps. 
Common potential parameters \cite{L15,L16}: $\kappa=0.8,A=1$,
1st Brillouin zone between $\pm \pi/2{\bf K}=\pm 0.7872$. Band edges: $\pm 0.6,\pm1$ (left plot), $\pm0.2428,\pm0.6472,\pm1.0290$ (right plot).}
\label{fig1}
\end{center}
\end{figure}

Why does the massless model (\ref{1.1}) with zero density of neutral and a finite density of charged fermions prefer the mean field of the massive model?
A look at the fermion dispersion relations belonging to both potentials gives a clue, see Fig.~\ref{fig1}. The mean field of the 
massless GN model features two gaps symmetric about 0, but no gap at zero energy. By contrast, the massive GN model gives rise to a mean field
with a third gap centered at $E=0$ (the reflection symmetry of both spectra is due to charge conjugation symmetry).
This is exactly what one would expect if each kind of fermion had its own gap at its Fermi surface, neutral fermions near zero
and charged fermions near some positive (for matter) or negative (for antimatter) Fermi energy. Hence we see here a first example of a system with two
Fermi surfaces and corresponding gaps in the context of the GN model family.

In order to further explore this kind of ``multiple Peierls instability", we have decided to study the NJL$_2$ model (\ref{1.2}) semi-classically along similar lines.
Here we shall again divide up the fermions into two groups, but allow for two arbitrary densities of the two fractions. As usual, we have to
start from a guess for the mean field and verify that the corresponding HF problem can be solved in a self-consistent way.
A natural candidate for the mean field in this situation is the twisted kink crystal \cite{L7,L8,L9}, with the additional bonus of being analytically tractable.
This will be taken as our starting point.

The paper is organized as follows. In Sec.~\ref{sec2}, we recall the most important properties of the original twisted kink crystal potential (with two
parameters) of Basar and Dunne \cite{L7,L8} and evaluate some integrals needed for fermion and energy densities later on. In Sec.~\ref{sec3}, we generalize
the potential to the four-parameter form used in Ref.~\cite{L9} by means of a scale transformation and a local chiral rotation. The results are then used in
Sec.~\ref{sec4} to solve the problem of two fermion components with two arbitrary densities within the HF approach. 
Sec.~\ref{sec5} contains a brief summary and our conclusions. Two appendices are devoted to a technicality of the regularization procedure (Appendix A) and 
the solution of the two-component problem assuming a homogeneous mean field (Appendix B).

\section{Basic twisted kink crystal of Basar and Dunne} 
\label{sec2}

Basar and Dunne have presented a candidate for a self-consistent mean field of the chiral Gross-Neveu model, using a particular ansatz  to solve the 
Eilenberger equation for the diagonal resolvent exactly. Here we start from the original two-parameter form of Ref.~\cite{L8}. The generalization to a four-parameter
potential introduced subsequently in \cite{L9} will be crucial below, but it is advantageous to first stick to the simpler form.  In the present section, we collect the most important 
ingredients and properties of the twisted kink crystal before embarking on applying it to a novel kind of physics problem. For more details and 
derivations, we refer to the original references \cite{L7,L8}.

The complex mean field for the NJL$_2$ model (``twisted kink crystal") has the rather complicated looking analytical expression
\begin{equation}
\hat{\Delta}(x) = -A \frac{\sigma(Ax+i{\bf K}' -i \theta/2)}{\sigma(Ax+i {\bf K}')\sigma(i \theta/2)} \exp\{iAx(-i \zeta(i \theta/2) + i \rm{ns}(i \theta/2))+ i  \zeta(i {\bf K}') \theta/2 \}
\label{2.1}
\end{equation}
with $A$ an amplitude factor given by
\begin{equation}
A  =   -2i\rm{sc}(i \theta/4) \rm{nd}(i \theta/4).
\label{2.2}
\end{equation}
Here and in the following, we suppress the 2nd and 3rd arguments of Weierstrass elliptic functions ($\zeta, \sigma, {\cal P}$) reading
\begin{eqnarray}
g_2 & = & \frac{4}{3} ( 1- \kappa^2 + \kappa^4 ),
\nonumber \\
g_3 & = & \frac{4}{27} (2-3 \kappa^2-3 \kappa^4+2 \kappa^6),
\label{2.3}
\end{eqnarray}
as well as the elliptic modulus $\kappa$ of Jacobi elliptic functions ($\rm{nc},\rm{nd},\rm{ns},\rm{sc}$) and complete elliptic integrals (${\bf E},{\bf K},{\bf K}'$).
We use units where the vacuum fermion mass is 1. The potential (\ref{2.1}) depends on two real parameters, $\kappa$ and $\theta$.
In the following chapter, we shall turn to the chirally twisted, rescaled form of $\hat{\Delta}$ from Ref.~\cite{L9},
\begin{equation}
\Delta(x) = \lambda e^{2iqx} \hat{\Delta}(\lambda x),
\label{2.4}
\end{equation}
which then depends on four real parameters.
Our notation will be such that  quantities derived from $\hat{\Delta}$ will be denoted by symbols with a hat, those derived from $\Delta$ by symbols without hat.  
The potential $\hat{\Delta}$ is quasi-periodic with two concomitant parameters, the period $L$ and a twist angle $\varphi$,
\begin{eqnarray}
\hat{\Delta}(x+L) & = &  e^{2i\varphi} \hat{\Delta}(x),
\nonumber \\
L & = & \frac{2 {\bf K}}{A},
\nonumber \\
\varphi & = & {\bf K}\{-i \zeta(i \theta/2) + i {\rm ns}(i \theta/2)\} - \zeta({\bf K}) \theta/2.
\label{2.5}
\end{eqnarray}
The Dirac fermion spectrum in this potential has a band structure with two gaps and band edges at
\begin{eqnarray}
\hat{E}_1 & = & -1,
\nonumber \\
\hat{E}_2 & =& -1+ 2 {\rm nc}^2(i \theta/4),
\nonumber \\
\hat{E}_3 & = & -1 + 2 {\rm nd}^2(i \theta/4),
\nonumber \\
\hat{E}_4 & = & 1,
\label{2.6}
\end{eqnarray}
by increasing energy. Thus, there is a bound band (or valence band) with $\hat{E_2}<E<\hat{E_3}$ inside the mass gap $-1 < E < 1$ known from the vacuum.
The negative (positive) continuum bands have $E<-1$ ($E>1$) like in the vacuum. The position of the bound band in the mass gap 
is governed by the parameter $\theta$, its width by $\theta$ and 
$\kappa$, see Ref.~\cite{L8}. A crucial quantity is the density of states inside the allowed bands,
\begin{equation}
\hat{\rho}(E) = \pm \frac{1}{2\pi} \frac{ 2 E^2- (\hat{E}_2+\hat{E}_3)E-(\hat{E}_2-\hat{E}_3)^2/4-1+Z}{\sqrt{(E^2-1)(E-\hat{E}_2)(E-\hat{E}_3)}},
\label{2.7}
\end{equation}
with the plus sign in the continuum bands, the minus sign in the bound band, and
\begin{equation}
Z = - A^2[{\cal P}(i \theta/2)+  \zeta({\bf K})/{\bf K}]
\label{2.8}
\end{equation}
equal to the spatial average of $|\hat{\Delta}|^2$.

In the following we need the fermion densities and energy densities per flavor of the negative energy continuum band 1 and the bound valence band 2.
For the fermion densities per flavor, one finds the simple expressions
\begin{eqnarray}
\hat{\cal N}_1 & = & \int_{-\Lambda/2}^{-1} dE \hat{\rho}(E) = \frac{\varphi}{\pi L} + \frac{\Lambda}{2\pi},
\nonumber \\
\hat{\cal N}_2 & = & \int_{\hat{E}_2}^{\hat{E}_3} dE  \hat{\rho}(E) = \frac{1}{L}.
\label{2.9}
\end{eqnarray}
$\Lambda$ is a cutoff parameter to be sent to infinity at the end of the calculation. 
In contrast to the momentum cutoff regularization used in previous works on the non-chiral Gross-Neveu model,
here we use an energy cutoff. The reason is explained in appendix A, where we verify that the two methods give the same
result for the simpler cases of the vacuum and the chiral spiral. 

The single particle energy densities per flavor can also be evaluated analytically but are significantly more involved. They can be expressed 
in terms of complete and incomplete elliptic integrals as
\begin{eqnarray}
\hat{\cal E}_1^{\rm sp} & = &  \int_{-\Lambda/2}^{-1} dE E \hat{\rho}(E) 
\nonumber \\
& = &  \frac{Z}{2\pi} \left( \ln (2+\hat{E}_3-\hat{E}_2) - \ln 2 - 1\right) + \frac{(4+3 \hat{E}_3^2-2\hat{E}_2\hat{E}_3-\hat{E}_2^2)}{16\pi}
\nonumber \\
& & - \frac{\left[(\hat{E}_2+\hat{E}_3)(\hat{E}_2\hat{E}_3+1)+4Z\right]}{4\pi \sqrt{(1-\hat{E}_2)(\hat{E}_3+1)}}F(p,s)
- \frac{(\hat{E}_2+\hat{E}_3)\sqrt{(1-\hat{E}_2)(\hat{E}_3+1)}}{4\pi} E(p,s) 
\nonumber \\
& & + \frac{Z(1+\hat{E}_2)}{\pi \sqrt{(1-\hat{E}_2)(\hat{E}_3+1)}}\Pi(p,r,s)  - \frac{\Lambda^2}{8\pi} - \frac{Z}{2\pi} \left( \ln \Lambda -1 \right),
\nonumber \\
\hat{\cal E}_2^{\rm sp} & = &  \int_{\hat{E}_2}^{\hat{E}_3} dE E   \hat{\rho}(E) 
\nonumber \\
& = &  \frac{\left[(\hat{E}_2+\hat{E}_3)(\hat{E}_2\hat{E}_3+1) +4Z\right]}{4\pi \sqrt{(1-\hat{E}_2)(\hat{E}_3+1)}}{\bf K}(s) + \frac{(\hat{E}_2+\hat{E}_3)\sqrt{(1-\hat{E}_2)(\hat{E}_3+1)}}{4\pi}{\bf E}(s)
\nonumber \\
& & - \frac{Z(1+ \hat{E}_2)}{\pi \sqrt{(1-\hat{E}_2)(\hat{E}_3+1)}}{\bf \Pi}(r,s),
\label{2.10}
\end{eqnarray}
where all  elliptic integrals are in MAPLE notation (see also Appendix A of Ref.~\cite{L17}) with the arguments 
\begin{eqnarray}
p & = & \sqrt{\frac{1-\hat{E}_2}{2}},
\nonumber \\
s & = & \sqrt{\frac{2(\hat{E}_3-\hat{E}_2)}{(1-\hat{E}_2)(\hat{E}_3+1)}},
\nonumber \\
r & = & \frac{\hat{E}_3-\hat{E}_2}{\hat{E}_3+1}=(ps)^2.
\label{2.11}
\end{eqnarray} 
The densities $\hat {\cal N}_{1,2}$ and $\hat{\cal E}_{1,2}^{\rm sp}$ are the main result of this section. They depend on the mean field $\hat{\Delta}$ through two real parameters,
$\kappa \in [0,1]$ and $\theta \in [0,4{\bf K}']$. 

\begin{figure}
\begin{center}
\epsfig{file=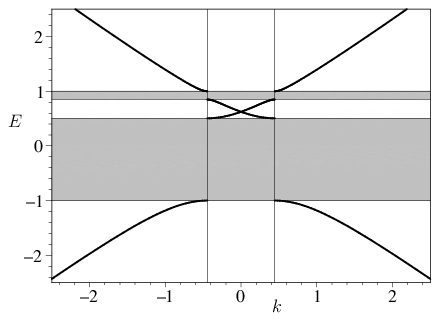,height=4cm,angle=0}\ \epsfig{file=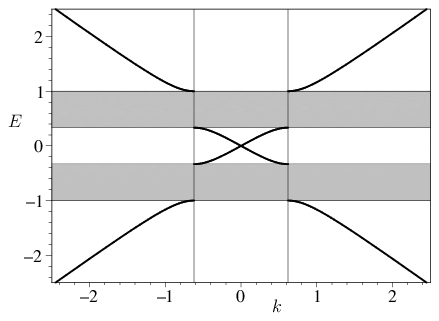,height=4cm,angle=0}\ \epsfig{file=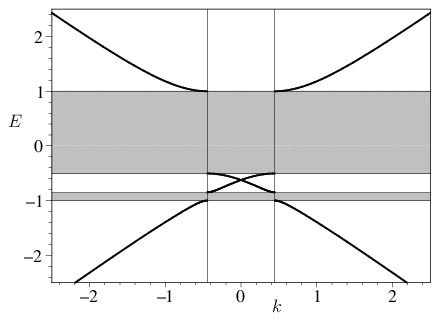,height=4cm,angle=0}
\caption{Examples of fermion dispersion relations for basic twisted kink potential $\hat{\Delta}$ \cite{L7,L8}. Shaded stripes: band gaps.
Parameters: $\kappa=0.5$, $\theta/\theta_{\rm max}=1/4, 1/2, 3/4$ from left to
right ($\theta_{\rm max} =4 {\bf K}'$). 1st Brillouin zone between  $\pm \pi/2L=\pm 0.4457,\pm 0.6212,\pm 0.4457$ from left to right.
Band edges: Continuum bands always bounded by $\pm 1$. The bound band in between has the edges $[0.5048,0.8480]$, $[-0.3333,0.3333]$, $[-0.8480,-0.5048]$
from left to right.}
\label{fig2}
\end{center}
\end{figure}

Let us now turn to the Dirac-HF or  BdG equation with potential $\hat{\Delta}(x)$. Using the representation of the $\gamma$
matrices
\begin{equation}
\gamma^0 = \sigma_1, \quad \gamma^1 = -i \sigma_2, \quad \gamma_5 = \gamma^0 \gamma^1 = \sigma_3,
\label{2.12}
\end{equation}
it assumes the form 
\begin{equation}
\left( \begin{array}{cc} -i   \partial_x & \hat{\Delta}(x) \\ \hat{\Delta}^*(x) & i \partial_x \end{array} \right) \psi = E \psi .
\label{2.13}
\end{equation}
The eigenspinors $\psi$ are also known exactly.  The energy and the
Bloch momentum are given parametrically by
\begin{eqnarray}
E(\alpha) & = & \frac{A}{2} \left\{ i \zeta(i\alpha-i\theta/4) -i \zeta(i\alpha+i\theta/4)+i\zeta(i \theta/2)+i {\rm ns}(i\theta/2) \right\},
\nonumber \\
k(\alpha) & = & - \frac{A}{2}\left\{ i \zeta(i\alpha-i \theta/4)+i \zeta(i \alpha+i \theta/4) + 2 \zeta({\bf K}) \alpha/{\bf K} \right\}.
\label{2.14}
\end{eqnarray}
The spectral parameter $\alpha$ is real for the continuum bands, ranging from $0...\theta/4$ in the negative energy band ($E<-1$) and from
${\bf K}'...\theta/4$ in the positive energy band ($E>1$). In the valence band ($\hat{E}_2 < E < \hat{E}_3$), $\alpha$ is
complex with $\rm{Im}\,\alpha = -{\bf K}, \rm{Re}\,\alpha=0...{\bf K}'$.
The Bloch momentum is only defined modulo $2 \pi/L$. By choosing appropriate branches and calibrating the momenta by an overall shift,
one can construct the fermion spectrum in the extended scheme (with Bloch momentum equal to 0 at the center of the Brillouin zone).
A few examples are shown in Fig.~\ref{fig2} for $\kappa=0.5$ and three different values of $\theta$, namely ${\bf K}'$, $2{\bf K}'$ and $3{\bf K}'$.
Vertical lines delimit the first Brillouin zone, horizontal lines belong to the band edges.
For $\theta=2{\bf K}'$ (central plot in Fig.~\ref{fig2}), the mean field $\hat{\Delta}$ is real and reduces to the scalar potential that solves 
the non-chiral, massless GN model. The other two plots are not symmetric about $E=0$ due to the lack of charge conjugation symmetry in 
the NJL$_2$ model. Notice that the plots corresponding to $\theta={\bf K}'$ and $\theta=3 {\bf K}'$ are related by a reflection
$E\to -E$. The corresponding mean fields are complex conjugates of each other, as is always the case if the $\kappa$-paramter is 
the same and the $\Theta$ parameters add up to $4 {\bf K}'$.
The bound band could be shrinked by increasing $\kappa$ or expanded by decreasing $\kappa$. As a function of $\theta$, it is widest at
$\theta=2{\bf K}'$, shrinking symmetrically to zero width at the end points $0$ and $4{\bf K}'$. 

\section{Rescaled, chirally twisted mean field} 
\label{sec3}

We actually need the generalized form of the mean field, $\Delta(x)$ of Eq.~(\ref{2.4}), proposed and studied in Ref.~\cite{L9}. It has four real parameters, 
$\kappa,\theta,\lambda,q$. As discussed in Ref.~\cite{L9}, the transformation (\ref{2.4}) reflects the scale and U(1) symmetries of the BdG equation
at the classical level, both broken quantum mechanically. Under such a transformation, a single particle state of energy $E$ and quasi-momentum $k$ 
goes over into a state with energy $\lambda E+q$ and quasi-momentum $\lambda k$. Again following Ref.~\cite{L9},
we express the fermion- and energy densities for completely filled lower continuum and bound bands for the potential $\Delta$ in terms
of the corresponding quantities for $\hat{\Delta}$. 
To evaluate the HF energy for given densities, four quantities will be needed later on,
\begin{eqnarray}
{\cal N}_1 & = & \int_{-\Lambda/2}^{E_1} dE \rho(E) ,
\nonumber \\
{\cal N}_2 & = & \int_{E_2}^{E_3} dE  \rho(E),
\nonumber \\
{\cal E}_1^{\rm sp} & = & \int_{-\Lambda/2}^{E_1} dE E \rho(E) ,
\nonumber \\
{\cal E}_2^{\rm sp} & = & \int_{E_2}^{E_3} dE  E \rho(E).
\label{3.1}
\end{eqnarray}
Rather than computing these integrals anew, we reduce them to the ones with a hat already available from the preceding section,
Eqs.~(\ref{2.9},\ref{2.10}). This is possible because the density of states, $\rho(E)$, can be expressed by $\hat{\rho}(E)$  via
\begin{equation}
\rho(E) = \hat{\rho} \left( \frac{E-q}{\lambda} \right),
\label{3.2}
\end{equation}
whereas the respective band edges are related by
\begin{equation}
E_i = \lambda   \hat{E}_i+q .
\label{3.3}
\end{equation}
It is important to use the same energy cutoff here and in Eqs.~(\ref{2.9},\ref{2.10}), as it should not depend on the specific potential $\Delta$.
Otherwise, one would miss the anomalous terms of the scaling and chiral symmetries. 
Together with the asymptotic form of $\hat{\rho}$,
\begin{equation}
\hat{\rho}(E) \approx \frac{1}{\pi} \left( 1 + \frac{Z}{2E^2} \right) \quad (E \gg 1),
\label{3.4}
\end{equation}
this is all it takes to express the four densities in (\ref{3.1}) in terms of the known hatted ones. 
Consider the fermion density in the lower continuum band first,
\begin{eqnarray}
{\cal N}_1 & = & \int_{-\Lambda/2}^{E_1} dE \rho(E) 
\nonumber \\
& = & \int_{-\Lambda/2}^{E_1} dE \hat{\rho}(E') \quad \left(E' = \frac{E-q}{\lambda}\right)
\nonumber \\
& = & \lambda \int_{-(\Lambda/2+q)/\lambda}^{-1} dE' \hat{\rho}(E')
\nonumber \\
& = &  \lambda \int_{-(\Lambda/2+q)/\lambda}^{-\Lambda/2} dE' \hat{\rho}(E') + \lambda \int_{-\Lambda/2}^{-1} dE' \hat{\rho}(E').
\label{3.5}
\end{eqnarray}
In the first term, we may now use the asymptotic form (\ref{3.4}) to find the analytical result. The second term 
is just $\lambda \hat{\cal N}_1$. Thus
\begin{equation}
{\cal N}_1 = \lambda \hat{\cal N}_1 + (1-\lambda)\frac{\Lambda}{2\pi}+ \frac{q}{\pi}.
\label{3.6}
\end{equation} 
This formula becomes more transparent upon introducing the finite, vacuum subtracted fermion densities
\begin{eqnarray}
\hat{\rho}_1 & = & \hat{\cal N}_1 - \frac{\Lambda}{2\pi},
\nonumber \\
\rho_1 & = & {\cal N}_1 - \frac{\Lambda}{2\pi}.
\label{3.7}
\end{eqnarray}
Eq.~(\ref{3.6}) simplifies to
\begin{equation}
\rho_1 = \lambda \hat{\rho}_1 + \frac{q}{\pi}.
\label{3.8}
\end{equation}
This is the familiar manifestation of the chiral anomaly --- fermion density is generated by a local chiral transformation, the very basis of
the chiral spiral. For the bound band, neither anomalous terms nor vacuum subtraction are necessary so that the corresponding equations
are simply
\begin{equation}
\hat{\rho}_2  =  \hat{\cal N}_2,  \quad \rho_2  =  {\cal N}_2, \quad \rho_2  =   \lambda \hat{\rho}_2 .
\label{3.9}
\end{equation}
Now let us turn to the single particle energy densities. In the lower continuum band
\begin{eqnarray}
{\cal E}_1^{\rm sp} & = & \int_{-\Lambda/2}^{E_1} dE E \rho(E) 
\nonumber \\
& = & \int_{-\Lambda/2}^{E_1} dE E \hat{\rho}(E') \quad \left(E' = \frac{E-q}{\lambda} \right)
\nonumber \\
& = & \lambda \int_{-(\Lambda/2+ q)/\lambda}^{-1} dE' (\lambda E'+q) \hat{\rho}(E')
\nonumber \\
& = &  \lambda^2 \int_{-(\Lambda/2+ q)/\lambda}^{-\Lambda/2} dE' E' \hat{\rho}(E') 
+ \lambda^2 \hat{\cal E}_1^{\rm sp} + \lambda q {\cal N}_1.
\label{3.10}
\end{eqnarray}
We have identified $\hat{\cal E}_1^{\rm sp}$ and ${\cal N}_1$ on the right-hand side.
The remaining integral can again be evaluated using the asymptotic form (\ref{3.4}) yielding
\begin{equation}
{\cal E}_1^{\rm sp} =  \lambda^2 \hat{\cal E}_1^{\rm sp} + \lambda q \hat{\cal N}_1 + (\lambda^2-1) \frac{\Lambda^2}{8 \pi} - \frac{\lambda q \Lambda}{2 \pi}+ \frac{q^2}{2\pi} 
+ \frac{Z \lambda^2 \ln \lambda}{2\pi} .
\label{3.11}
\end{equation}
To eliminate the spurious cutoff dependence, the result is best expressed via vacuum subtracted
quantities (eliminating the quadratic divergence) and after adding the double counting correction (eliminating the log divergence). This yields the physical HF energy density
\begin{eqnarray}
\hat{\cal E}_1 & = & \hat{\cal E}_1^{\rm sp} + \frac{\Lambda^2}{8\pi} + \hat{\cal E}_{\rm dc},
\nonumber \\
{\cal E}_1 & = & {\cal E}_1^{\rm sp} + \frac{\Lambda^2}{8\pi} +  {\cal E}_{\rm dc},
\label{3.12}
\end{eqnarray}
where
\begin{equation}
\hat{\cal E}_{\rm dc}  =  {\cal E}_{\rm dc} /\lambda^2 =  \frac{Z}{2\pi} \left( \ln \Lambda -1 \right).
\label{3.13}
\end{equation}
The subleading term on the right-hand side is due to our use of an energy cutoff, see appendix A.
Adding ${\cal E}_{\rm dc}$ on both sides of (\ref{3.11}) and rearranging the various terms, we finally arrive at
\begin{equation}
{\cal E}_1 = \lambda^2 \hat{\cal E}_1 + \lambda q \hat{\rho}_1 + \frac{q^2}{2\pi} + \frac{Z \lambda^2 \ln \lambda}{2\pi}.
\label{3.14}
\end{equation}
The corresponding calculation for the valence band is free of divergencies, yielding the simpler scaling relation
\begin{equation}
{\cal E}_2 = \lambda^2 \hat{\cal E}_2 + \lambda q \hat{\rho}_2 .
\label{3.15}
\end{equation}
Eqs.~(\ref{3.8},\ref{3.9},\ref{3.14},\ref{3.15}) express how the transition from $\hat{\Delta}$ to $\Delta$ manifests itself 
in the relevant observables. These depend on all four parameters of the mean field. However, the dependence on $\lambda$ and $q$
is now simple and explicit, whereas the dependence on $\theta$ and $\kappa$ through $\hat{\cal N}_i, \hat{\cal E}_i$ is rather
complicated, see Eqs.~(\ref{2.9},\ref{2.10}).

We are now equipped with all necessary tools to attack a real physics problem. This is the subject of next section.

\section{NJL$_2$ model with two different Fermion densities} 
\label{sec4}

We would like to solve the following problem within the NJL$_2$ model. Let us divide the $N$ fermion flavors up into two subsets denoted as ``type $A$" (fraction ($1-\nu$)) and ``type $B$" (fraction $\nu$).
Since flavor is conserved, we are free to prescribe two different densities for these two fermion species, $\rho_A$ and $\rho_B$. What is the HF ground state of such a system?
As pointed out in the introduction, we anticipate that two different Fermi surfaces will lead to opening two distinct gaps in the spectrum of the HF potential. If this is true, the twisted chiral spiral is
a natural trial potential to search for the true HF ground state. Indeed, it features a band spectrum with two gaps which can be moved to any desired energy, thanks to the rescaling parameters $\lambda,q$.
We expect fermions of type $A$ to completely fill band 1 and fermions of type $B$  to completely fill bands 1 and  2 in the ground state. Thus, the relationship between their fermion and energy densities and the
quantities computed above is 
\begin{eqnarray}
\rho_A &  = & (1-\nu) \rho_1,
\nonumber \\
\rho_B & = & \nu (\rho_1 + \rho_2),
\nonumber \\
{\cal E}_A & = & (1-\nu) {\cal E}_1,
\nonumber \\
{\cal E}_B & = & \nu ({\cal E}_1+{\cal E}_2),
\label{4.1}
\end{eqnarray}
whereas the total fermion density and energy density are given by
\begin{eqnarray}
\rho & = & \rho_A+ \rho_B = \rho_1 + \nu \rho_2,
\nonumber \\
{\cal E} & = & {\cal E}_A + {\cal E}_B = {\cal E}_1 + \nu {\cal E}_2.
\label{4.2}
\end{eqnarray}
Prescribing $\rho_A,\rho_B$ for a given fraction $\nu$ of type $B$ fermions is obviously equivalent to prescribing $\rho_1,\rho_2$. We then have to minimize the total 
energy subject to the constraints for both densities with respect to the four parameters in $\Delta(x)$. 
As we proceed to show now, three out of these four parameters can be eliminated and the problem reduces to finding the minimum of a function
of one variable only.

Suppose we prescribe $\nu,\rho_A$  and $\rho_B$ and determine $\rho_1,\rho_2$. Here, $\rho_1$ can have any sign, whereas we shall assume that $\rho_2>0$
to ensure that the Fermi energies are in the right order. We can then solve Eqs.~(\ref{3.8},\ref{3.9}) for $\lambda$ and $q$,
\begin{eqnarray}
\lambda & = & \frac{\rho_2}{\hat{\rho}_2}=L \rho_2,
\nonumber \\
\frac{q}{\pi} & = & \frac{\rho_1 \hat{\rho}_2 - \rho_2 \hat{\rho}_1}{\hat{\rho}_2} =\rho_1-\frac{\varphi}{\pi} \rho_2.
\label{4.3}
\end{eqnarray}
This eliminates two out of the four parameters of $\Delta(x)$  right away.
According to (\ref{3.14},\ref{3.15},\ref{4.2}), the total energy density to be minimized is 
\begin{equation}
{\cal E} = \lambda^2 \hat{\cal E} + \lambda q \hat{\rho} + \frac{q^2}{2\pi} + \frac{\lambda^2 Z \ln \lambda}{2\pi},
\label{4.4}
\end{equation}
where $\hat{\rho}, \hat{\cal E}$ are defined in analogy to (\ref{4.2}).
It is worthwhile inserting $\lambda,q$ from (\ref{4.3}) and decompose $\hat{\rho}$ again as in (\ref{4.2}) with the result
\begin{equation}
{\cal E}- \frac{\pi \rho_1}{2} \left( \rho_1+ 2 \nu \rho_2 \right) = \frac{\rho_2^2}{\hat{\rho}_2^2} \left( \hat{\cal E}- \frac{\pi\hat{\rho}_1}{2} \left( \hat{\rho}_1+ 2 \nu \hat{\rho}_2 \right) 
+ \frac{Z}{2\pi} \ln  \frac{\rho_2}{\hat{\rho}_2} \right).
\label{4.5}
\end{equation}
Remarkably, $\rho_1$ only enters in a ($\kappa,\theta$) independent term so that the minimization of ${\cal E}$ needs to be done at $\rho_1=0$ only. The results for finite $\rho_1$ are
then obtained for free,
\begin{equation}
{\cal E}(\nu,\rho_1,\rho_2) = {\cal E}(\nu, 0 ,\rho_2) +  \frac{\pi \rho_1}{2} \left( \rho_1+ 2 \nu \rho_2 \right) .
\label{4.6}
\end{equation}
From now on, we set $\rho_1=0$.

To proceed, we take the expression for ${\cal E}$, Eq.~(\ref{4.5}), and insert $\hat{\cal E},\hat{\rho}$, using the analytic formulas (\ref{2.9}-\ref{2.11}). In the spirit of a variational calculation, we 
could then minimize the result with respect to $\kappa,\theta$. However we are interested in true HF solutions. This raises the issue of self-consistency of the mean field, i.e., the condition 
\begin{equation}
\Delta = -   Ng^2 \sum_{\rm occ}2  \psi_1 \psi_2^* .
\label{4.7}
\end{equation}
In order to verify this equation, we need the eigenspinors in addition to what has already been discussed above. These are also known in closed form. Actually, the
contribution from a single eigenstate to the sum on the right-hand side of (\ref{4.8}) has already been given in Ref.~\cite{L8}, at least for the basic potential $\hat{\Delta}$.
Since the single particle HF spinors transform as
\begin{equation}
\psi(x) = (\lambda)^{-1/2} e^{iq\gamma_5 x} \hat{\psi}(\lambda x)
\label{4.8}
\end{equation}
under the transformation (\ref{2.4}), we can immediately generalize this result to the rescaled case, finding
\begin{equation}
2 \psi_1(x) \psi_2^{*}(x) = \pm \frac{1}{dE/d\alpha} \frac{1}{\lambda A} ( b(E) \Delta(x) - i \partial_x \Delta(x))
\label{4.9}
\end{equation}
with 
\begin{eqnarray}
b(E) & = & 2E- (E1+E_2+E_3+E_4),
\nonumber \\
\frac{dE}{d\alpha} & = &  \frac{2}{A} \sqrt{(E-E_1)(E-E_2)(E-E_3)(E-E4)},
\label{4.10}
\end{eqnarray}
and the plus sign for the continuum bands, the minus sign for the bound band.
What is new as compared to Refs.~\cite{L8,L9} is the way in which we occupy the single particle states. In the present case, the sum over occupied states has to be
interpreted as integral over band 1 (the lower continuum band) plus $\nu$ times the integral over band 2 (the bound band).
Owing to the structure of Eq.~(\ref{4.9}), the self-consistency condition is equivalent to two
$x$-independent relations,
\begin{eqnarray}
\left(\int_{-\Lambda/2}^{E_1}dE - \nu \int_{E_2}^{E_3}dE \right) \frac{1}{dE/d\alpha} \frac{1}{\lambda A}  b(E) + \frac{\pi}{Ng^2} & = & 0,
\label{4.11} \\
\left(\int_{-\Lambda/2}^{E_1}dE - \nu \int_{E_2}^{E_3}dE \right) \frac{1}{dE/d\alpha} & = & 0.
\label{4.12}
\end{eqnarray}
The minus sign between the contributions from the two bands is due to the $\pm$ sign in Eq.~(\ref{4.9}). Eq.~(\ref{4.11}) is still somewhat symbolic due to the logarithmic divergence
of both terms. We will return to it later.

Consider Eq.~(\ref{4.12}) first, i.e., the condition that the terms proportional to the derivative of $\Delta(x)$ cancel. As already noted in Ref.~\cite{L8}, the left-hand side of the 2nd equation
can be evaluated trivially, being just an integral over the spectral parameter $\alpha$. Taking into account the $\alpha$-values at the band edges, find
\begin{eqnarray}
\int_{0}^{\theta/4} d \alpha & = &  \theta/4 \quad ({\rm band\ 1}) ,
\nonumber \\
\int_{-i{\bf K}}^{-i{\bf K}+{\bf K}'}d \alpha & = & {\bf K}' \quad ({\rm band\ 2}).
\label{4.13}
\end{eqnarray}
Hence the second self-consistency condition, Eq.~(\ref{4.12}), yields an explicit relation between $\theta$ and $\kappa$ for given $\nu$,
\begin{equation}
\theta = 4 \nu {\bf K}'.
\label{4.14}
\end{equation}
Eliminating $\theta$ in this manner leaves us with a single real parameter, the elliptic modulus $\kappa$. The easiest way to proceed is to minimize the energy density
with respect to $\kappa$ and come back to the other part of the self-consistency condition, Eq.~(\ref{4.11}), afterwards. 

We now turn to the results of our calculation. We have chosen three values for the fraction $\nu$ of the $N$ fermion flavors with the higher density, $\nu =$ 1/4, 1/2, and 3/4. These are the same values
for which we have shown the dispersion relation before rescaling, see Fig.~\ref{fig2}. For the same value of $\kappa$, these figures could also be used for the rescaled potential except for
a relabeling of the axes ($k \to \lambda k, E \to \lambda E+q$). All calculations were done for $\rho_1=0$, since the results for any finite $\rho_1$ can be obtained using Eq.~(\ref{4.6}).

\begin{figure}
\begin{center}
\epsfig{file=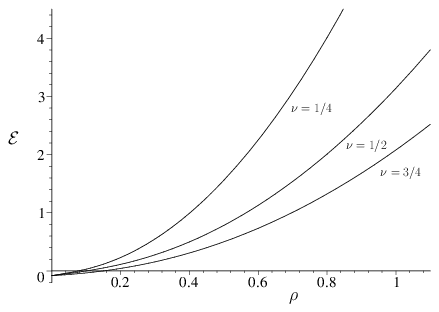,height=6cm,angle=0}
\caption{Energy density vs. fermion density from two-component HF calculation based on the twisted kink crystal. Densities: $\rho_1=0$ (type $A$ fermions, fraction $1-\nu$),
$\rho_2= \rho/\nu$ (type $B$ fermions, fraction $\nu$) with $\rho$ on the horizontal axis. Three values of $\nu$ are shown. On this scale, results are practically indistinguishable from 
those based on a homogeneous mean field, see Fig.~\ref{fig11}, Appendix B.}
\label{fig3}
\end{center}
\end{figure}

\begin{figure}
\begin{center}
\epsfig{file=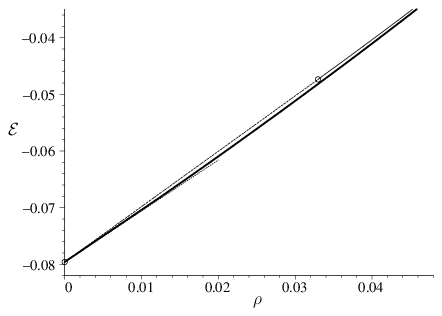,height=5cm,angle=0}\ \ \epsfig{file=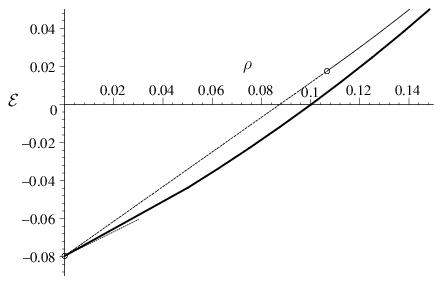,height=5cm,angle=0}
\caption{Zooming in onto the small $\rho$ region of Fig.~\ref{fig3}. Fat curves: Details of curves labeled $\nu=1/4$ (left plot) and $\nu=1/2$ (right plot).
The thin curves are from the homogeneous calculation, Appendix B. The circles correspond to the critical point where a first order
phase transition occurs. These points are joined by a straight (dashed) line to the vacuum point at $\rho=0$, the mixed phase. The full calculation is lower in energy
and does not show any phase transition.}
\label{fig4}
\end{center}
\end{figure}

\begin{figure}
\begin{center}
\epsfig{file=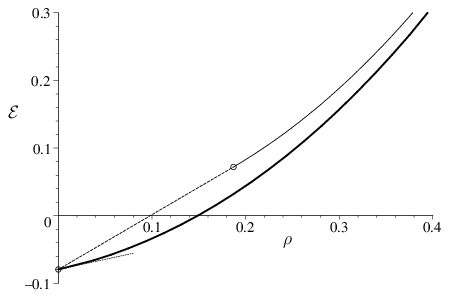,height=5cm,angle=0}\ \ \epsfig{file=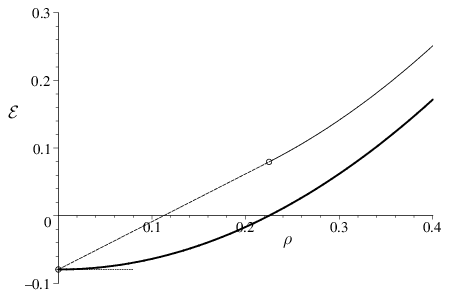,height=5cm,angle=0}
\caption{Like Fig.~\ref{fig4}, but for $\nu=3/4$ (left plot) and $\nu=1$ (right plot).  The $\nu=1$ curve is the result from the chiral spiral, not shown in Fig.~\ref{fig3}.
The short dotted lines tangential to the full curves at $\rho=0$ (also drawn in Fig.~\ref{fig3}) show that the slope is consistent with the mass of a single 
twisted kink.}  
\label{fig5}
\end{center}
\end{figure}

The central result is a plot of the energy density vs. fermion density ($\rho=\nu \rho_2$), Fig.~\ref{fig3}, for three values of $\nu$. All three curves start form the renormalized vacuum energy
density, $-1/4\pi$, at zero density and quickly approach the value for free, massless fermions, ${\cal E}= \pi \rho^2/2 \nu$, at high densities. 
As a matter of fact, at this scale one can hardly see the difference between this calculation and an alternative one based on homogeneous
condensates (see Appendix B, Fig.~\ref{fig11}). To highlight the differences between both approaches indicative of the Peierls effect, we therefore 
zoom into the small density region in Figs.~\ref{fig4}, \ref{fig5}. The fat, solid curves are the result of the full calculation using the twisted kink crystal potential.
We have also included the results for $\nu=1$ corresponding to the standard chiral spiral in Fig.~\ref{fig5}, where only a single fermion component is left.
The thin lines refer to the homogeneous calculation. As discussed in Appendix B, they exhibit a first order phase transition at some critical density (marked by a 
circle) where the fermion mass drops from the vacuum value 1 to a smaller, finite value. Below this point, one is in a mixed phase characterized by a linear behaviour
of ${\cal E}$ vs. $\rho$ (thin dashed line). This line is tangential to the thin curve beginning at the critical density, as expected from a Maxwell construction.
We have also drawn a dotted line segment corresponding to the expected behavior of the full calculation near $\rho=0$,
\begin{equation}
{\cal E}-{\cal E}_{\rm vac}  \approx \frac{1}{\nu} M_{\rm kink}(\nu) \rho, \quad M_{\rm kink}(\nu) = \frac{\sin \pi \nu}{\pi}.
\label{4.15}
\end{equation}
The mass of the single twisted kink with occupation fraction $\nu$ is taken from the original reference \cite{L18}. The fact that it yields the correct low density behavior 
of the crystal calculation is a useful independent test of our results. 
The discrepancy between the full calculation (with gapped spectrum) and the homogeneous one diminishes with decreasing $\nu$, becoming almost negligible at $\nu=1/4$. 
Nevertheless, the accuracy of the present calculation is sufficient to demonstrate that the full calculation always yields a lower result, proving that a kind
of ``double Peierls effect" is at work.

\begin{figure}
\begin{center}
\epsfig{file=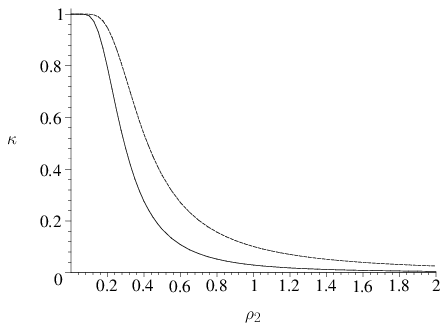,height=5cm,angle=0}\ \ \epsfig{file=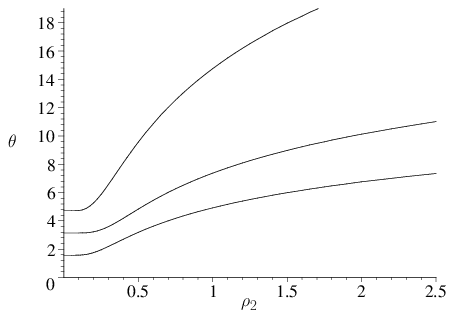,height=5cm,angle=0}
\caption{Potential parameters $\kappa$ (left plot) and $\theta$ (right plot) determined in the two-component HF calculation, vs. $\rho_2$. The wide curve
in the left plot is from $\nu=1/2$, the narrow curve from $ \nu=1/4,3/4$. In the right plot, the three curves correspond to $\nu=1/4,
1/2, 3/4$ from bottom to top.}
\label{fig6}
\end{center}
\end{figure}

\begin{figure}
\begin{center}
\epsfig{file=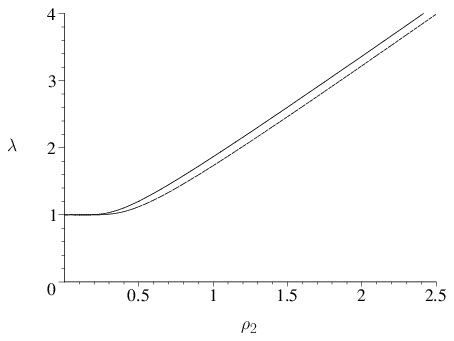,height=5cm,angle=0}\ \ \epsfig{file=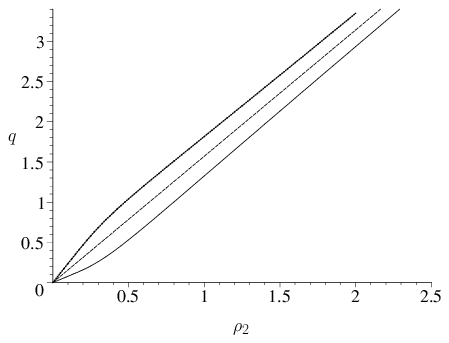,height=5cm,angle=0}
\caption{Potential parameters $\lambda$ (left plot) and $q$ (right plot) determined in the two-component HF calculation, vs. $\rho_2$. Left plot: lower curve
is for $\nu=1/2$, upper curve for $\nu=1/4,3/4$. Right plot: $\nu=1/4,1/2,3/4$ from bottom to top.}              
\label{fig7}
\end{center}
\end{figure}

\begin{figure}
\begin{center}
\epsfig{file=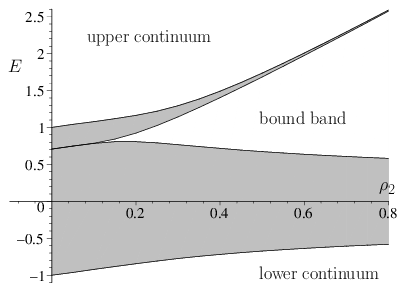,height=5cm,angle=0}\ \ \epsfig{file=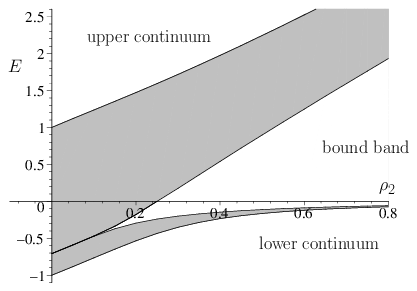,height=5cm,angle=0}
\caption{Evolution of band edges against $\rho_2$, for self-consistent two-component HF calculation. Shaded areas: Gaps. 
Left plot: $\nu=1/4$, right plot: $\nu=3/4$. The endpoint of the bound band at $\rho=0$ agrees with the position of the bound state
in the single twisted kink, $E_0=\cos \nu\pi$.}
\label{fig8}
\end{center}
\end{figure}

When generating these results, we had to eliminate analytically the potential parameters $\lambda,q,\theta$ and minimize numerically with respect 
to $\kappa$. We show all four parameters in Figs.~\ref{fig6} and \ref{fig7} against $\rho_2$. Fig.~\ref{fig6} exhibits the parameters $\kappa,\theta$ intrinsic to the original twisted kink crystal $\hat{\Delta}$
of Ref.~\cite{L8}, whereas Fig.~\ref{fig7} shows the extra parameters $\lambda,q$ introduced when rescaling and chirally twisting this potential to arrive at the
full potential $\Delta$. From the point of view of the Peierls instability, it is interesting to ask about the band structure of the self-consistent potential
belonging to a certain density $\rho_2$. The band edges can be computed from the four parameters plotted in Figs.~\ref{fig6},\ref{fig7}. 
In Fig.~\ref{fig8} we show the results for the gap structure at $\nu=1/4$ (left) and $ \nu=3/4$ (right). The shaded regions are the gaps. The difference between these two plots is quite striking. For $\nu=1/4$,
the upper gap becomes negligible at high densities where the picture resembles the vacuum with a mass gap around zero. For $\nu=3/4$, the lower gap disappears at high densities, whereas 
the upper one moves upwards, reminiscent of the chiral spiral ($\nu=1$). The valence band becomes narrow at low energies, ending exactly at the point where the single 
twisted kink has its bound state, $E_0=\cos \pi \nu = \pm 1/\sqrt{2}$.   

\begin{figure}
\begin{center}
\epsfig{file=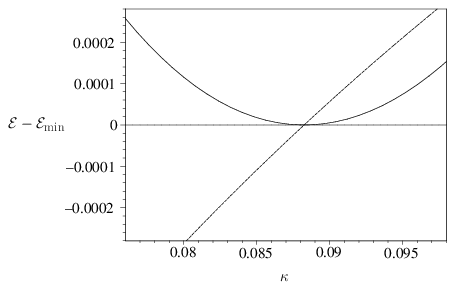,height=5cm,angle=0}
\caption{Example of self-consistency test ($\nu=3/4,\rho_2=0.65$). The parabolic curve is the energy density minus its minimum value, showing a minimum
at $\kappa=0.088$. Only a small region around the minimum is shown to enhance the sensitivity of the test. The linear curve is the 
self-consistency equation, Eq,~(\ref{4.11}), with $\pi/Ng^2$ replaced by $\ln \Lambda$,  in arbitrary units. The fact that it crosses zero right at the minimum
confirms the self-consistency of the HF solution.}
\label{fig9}
\end{center}
\end{figure}

Let us now come back to the first self-consistency relation, Eq.~(\ref{4.11}),  to check whether our solution is a valid HF solution or just variational.
The integral has a logarithmic divergence which is cancelled against the $\ln \Lambda$ in the inverse coupling constant, so that the result is finite. 
We have checked that this quantity vanishes exactly at the value of $\kappa$ found by minimization of the energy density, see Fig.~\ref{fig9} for an example. 
The plot is shown at a high resolution to underline the accuracy of the test.

\section{Summary and conclusions} 
\label{sec5}

In this work we have studied the NJL$_2$ model at finite density in a novel direction. Motivated by the recent work on the finite $N$ GN model \cite{L14}, 
we introduce two fermion densities for two subsets of the $N$ flavors. In the case of a single baryon, the corresponding study would be the 
step from the massless baryon with fermion number $N$ (where the mean field traces out a full turn around the chiral circle) to the twisted kink
of Shei \cite{L18} with partially filled valence level and fermion number $<N$. The difference between the two kinds of bound states is due to partially filled single particle states.
Likewise, in the case of matter at finite density, the new aspect is to allow for a partially filled valence band. Partial filling here does not mean that all
states up to a Fermi energy inside the band are completely filled, but that each state within the bound band is filled with the same fraction $\nu$ only.

The analogue to the step from the massless baryon to the twisted kink in the matter case is the step from the chiral spiral to the twisted kink
crystal. Indeed, the chiral spiral and the massless baryon differ only in the pitch of the helical mean field. Similarly, the twisted kink crystal 
and the single twisted kink are closely related: The twisted kink crystal mean field can be obtained from a periodic array of twisted kinks and
reduces to the isolated twisted kink in the limit $\kappa\to 1$ \cite{L8}. We have tried to make this relationship concrete by a full HF calculation
of the ground state energy of a system with two different fermion densities.

Aside from better understanding the physics of the twisted kink crystal, we have shown that relativistic quantum field theory models 
illustrate nicely the phenomenon one might call ``multiple Peierls instability". A system with several distinct Fermi surfaces may open
several gaps, one at each Fermi energy, giving rise to mean fields of increasing complexity. A first example was already given in Ref.~\cite{L14}
in the context of the GN model with discrete chiral symmetry. We can now contribute another example with continuous chiral symmetry. 
Actually, the elliptic mean fields that have
been found useful in this context are often part of a hierarchy of finite gap potentials with ever increasing number of gaps. This may give a
hint as to what kind of systems might be described by such more complicated fields. The most efficient strategy to find self-consistent fields
may be to start from the band structure of a certain potential and think about how one would fill the allowed bands most efficiently, then try to
find a field theory model that can accomodate this occupation scheme.
 
Our last remark is about equilibrium thermodynamics and finite temperature, a subject not touched in the present work. As far as the standard NJL$_2$ model
with a single fermion species only is concerned, the chiral spiral mean field is again preferred over the general twisted kink crystal \cite{L9}.
A result of previous studies is the fact that the self-consistent potential at zero and finite temperature often have the same functional form. 
This raises the question about the fate of the two-component NJL$_2$ model studied in the present work at finite temperature.
To answer this question, one would have to introduce again two kinds of fermions and attribute two different chemical potentials to them.
If the grand canonical potential for the one-component system is
\begin{equation}
\Psi^{(1)}(T,\mu) = - \frac{1}{\beta} \int dE \rho(E) \ln \left( 1+ e^{-\beta(E-\mu)} \right) + \frac{1}{2Ng^2} \frac{1}{L} \int_0^L dx |\Delta(x)|^2,
\label{5.1}
\end{equation} 
the two-component system with fractions ($1-\nu,\nu$) has the effective potential
\begin{equation}
\Psi^{(2)}(T,\nu,\mu_1,\mu_2) = (1-\nu) \Psi^{(1)}(T,\mu_1) + \nu \Psi^{(1)}(T,\mu_2).
\label{5.2}
\end{equation}
This has to be minimized  with respect to all parameters of the potential at each temperature and pair of chemical potentials. The self-consistency condition gets replaced by
\begin{equation}
\Delta(x) = - Ng^2 \int dE \rho(E)\left\{ \frac{1-\nu}{1+e^{\beta(E-\mu_1)}}+ \frac{\nu}{1+e^{\beta(E-\mu_2)}}\right\}  \bar{\psi}_E\psi_E. 
\label{5.3}
\end{equation}
Although this is likely to be a tedious task, it would be interesting to study the phase diagram of this system in ($T,\mu_1,\mu_2$) space.
If one uses the twisted kink crystal potential, all building blocks for such a computation are already available.

\section*{Appendix A: Energy vs. momentum cutoff for vacuum and chiral spiral} 
\label{sec6}

We have to face the problem how to regularize and renormalize the system. Due to the presence of twist and the axial anomaly, this is quite delicate.
In this work, we have decided to adopt a cutoff in energy rather than the usual momentum cutoff. To explain the reason and check that it works, we 
briefly discuss the vacuum and the chiral spiral in the NJL$_2$ model. We use a notation emphasizing the similarity with the procedure followed in
the main text, but do not set $m=1$ in the vacuum. 

Consider first the HF vacuum. The density of states is 
\begin{equation}
\hat{\rho}(E) = \frac{1}{\pi} \left|\frac{dk}{dE}\right| = \frac{|E|}{\sqrt{E^2-m^2}}.
\label{A1}
\end{equation}
The fermion density and energy density of the filled Dirac sea are given by 
\begin{eqnarray}
\hat{\cal N} & = & \int_{-\Lambda/2}^{-m} dE \hat{\rho}(E) = \frac{\Lambda}{2\pi},
\nonumber \\
\hat{\cal E} & = & \int_{-\Lambda/2}^{-m} dE E \hat{\rho}(E) = - \frac{\Lambda^2}{8\pi}+ \frac{m^2}{4\pi} - \frac{m^2}{2\pi} \ln\frac{\Lambda}{m} +  \frac{m^2}{2 Ng^2}.
\label{A2}
\end{eqnarray}
We have added the double counting correction to the single particle energy density. Minimizing $\hat{\cal E}$ with respect to $m$ yields the gap equation
in the form
\begin{equation}
m \left\{ 1-  \frac{Ng^2}{\pi} \left(\ln\frac{\Lambda}{m}-1\right) \right\} = 0.
\label{A3}
\end{equation}
In the chirally broken phase present at $T=0$, $m\neq 0$ and we get the condition
\begin{equation}
\frac{\pi}{Ng^2} = \ln\frac{\Lambda}{m} -1.
\label{A4}
\end{equation}
In the conventional momentum cutoff scheme, the $-1$ is absent on the right-hand side. Nevertheless, re-inserting (\ref{A4}) into the formula for ${\cal E}$ yields
exactly the same result as in momentum regularization,
\begin{equation}
\hat{\cal E} = - \frac{\Lambda^2}{8 \pi} - \frac{m^2}{4\pi}.
\label{A5}
\end{equation} 
Now consider the self-consistency condition
\begin{equation}
m = - Ng^2 \langle \bar{\psi}\psi \rangle = - Ng^2 \int_{-\Lambda/2}^{-m} dE \hat{\rho}(E) \frac{m}{E} = Ng^2 \frac{m}{\pi} \ln \frac{\Lambda}{m}.
\label{A6}
\end{equation}
It is telling us that
\begin{equation}
\frac{\pi}{Ng^2} = \ln \frac{\Lambda}{m}.
\label{A7}
\end{equation}
There is a discrepancy with Eq.~(\ref{A4}) in the subleading term, not encountered in the momentum cutoff scheme. 

We now turn to the chiral spiral, the solution of the NJL$_2$ model at finite density. The transition from the vacuum to the chiral spiral is reminiscent of
the transition from $\hat{\Delta}$ to $\Delta$ in the main text. A unitary transformation with $e^{iq\gamma_5 x}$
turns the scalar potential $m$ into the scalar--pseudoscalar mean field $m e^{2iqx}$, shifting the fermion spectrum rigidly upward by $q$. The density
of states becomes
\begin{equation}
\rho(E)=\frac{1}{\pi}\frac{|E-q|}{\sqrt{(E-q)^2-m^2}}= \hat{\rho}(E-q).
\label{A8}
\end{equation}
The relevant fermion and energy densities are 
\begin{eqnarray}
{\cal N} & = & \int_{-\Lambda/2}^{-m+q} dE \rho(E) ,
\nonumber \\
{\cal E} & = & \int_{-\Lambda/2}^{-m+q} dE E \rho(E) +  \frac{m^2}{2 Ng^2}.
\label{A9}
\end{eqnarray}
As in the main text, we can reduce expressions (\ref{A9}) to the ones with a hat computed before and related to the vacuum, 
\begin{eqnarray}
{\cal N} & = & \hat{\cal N} + \int_{- \Lambda/2-q}^{-\Lambda/2} dE' \hat{\rho}(E'),
\nonumber \\
{\cal E} & = & \hat{\cal E} + q {\cal N} + \int_{- \Lambda/2-q}^{-\Lambda/2} dE' E'\hat{\rho}(E') +  \frac{m^2}{2 Ng^2}.
\label{A10}
\end{eqnarray}
The extra terms can be evaluated readily by using the asymptotic form of $\hat{\rho}(E)$ for large $|E|$,
\begin{equation}
\hat{\rho}(E) \approx \frac{1}{\pi} \left( 1 + \frac{m^2}{2E^2} \right),
\label{A11}
\end{equation}
with the final result
\begin{eqnarray}
{\cal N} & = & \frac{\Lambda}{2\pi}+ \frac{q}{\pi},
\nonumber \\
{\cal E} & = & - \frac{\Lambda^2}{8\pi} - \frac{m^2}{4\pi}  + \frac{q^2}{2\pi}.
\label{A12}
\end{eqnarray}
After subtracting the divergent contributions from the Dirac sea, this reproduces the known results from the chiral spiral.
The self-consistency condition follows from that of the vacuum, the local chiral rotation being responsible for the 
inhomogeneity of the mean field. 

Now we come to the reason why we prefer the energy cutoff to momentum cutoff in the present work, despite
the unusual apparent (sub-leading) conflict between gap equation and self-consistency condition mentioned above.
If one regularizes the chiral spiral with momentum cutoff, one has to change the cutoff value from $\Lambda/2$ 
to $ \Lambda/2 +q$. Otherwise, it is not possible to account for the chiral anomaly. This recipe depends on
the special form of the mean field. It is not clear to us how to generalize it to the twisted kink crystal, where 
the anomaly enters in a more complicated fashion.

\section*{Appendix B: Homogeneous solution for system with two densities} 
\label{sec7}

To assess whether the Peierls effect is really favored, here we study the two-component HF problem with a homogeneous mean field. We assume that only a fraction $\nu$ of the 
$N$ fermion species has a finite density. Since we shall encounter a first order phase transition and a mixed phase, it is advantageous to work with a chemical 
potential as Lagrange multiplier. The grand canonical potential at zero temperature reads
\begin{equation}
{\cal V}_{\rm eff} = \frac{m^2 \ln (m^2)}{4 \pi} - \frac{m^2}{4\pi} + \theta(\mu-m) \frac{\nu}{2\pi} \left\{ m^2 \ln \left(\frac{\mu+\sqrt{\mu^2-m^2}}{m}\right) - \mu \sqrt{\mu^2-m^2} \right\}.
\label{B1}
\end{equation}
The standard result for one component is recovered at $\nu=1$ \cite{L19}. There, the minimum of ${\cal V}_{\rm eff}$ with respect to $m$ stays at the vacuum value up to the critical 
chemical potential $\mu_{\rm crit} = 1/\sqrt{2}$, then it drops to 0. Chiral symmetry gets restored in a first order phase transition. For $\nu<1$, we find a different behavior.
The minimum is again at the vacuum value, until it jumps to some smaller but finite value at a critical chemical potential. This is again a first order transition, but chiral symmetry does not get 
restored, similar to what is found in the  massive GN model. For larger $\mu$, the mass decreases monotonically without ever reaching the $m=0$ axis.
This can be seen more quantitatively as follows. 
Starting from 
\begin{equation}
\frac{\partial {\cal V}_{\rm eff}}{\partial m} = 0,
\label{B2}
\end{equation}
one finds that $m=0$ is always a possible solution. The other solution has to satisfy
\begin{equation}
\mu + \sqrt{\mu^2-m^2} = m^{(\nu-1)/\nu}.
\label{B3}
\end{equation}
This cannot be solved in closed analytical form for $m$, given $\mu$. Using Eq.~(\ref{B3}) to eliminate $\sqrt{\mu^2-m^2}$ from ${\cal V}_{\rm eff}$, we can express the value of 
${\cal V}_{\rm eff}$ at the minimum as
\begin{equation}
\left. {\cal V}_{\rm eff}\right|_{\rm min} = - \frac{1}{4\pi} \left(m^2+2\nu \mu \sqrt{\mu^2-m^2}\right).
\label{B4}
\end{equation}
The phase transition happens at the chemical potential where the right-hand side equals the vacuum energy density $-1/(4\pi)$. This yields the condition
obtained by plugging the solution of $\left. {\cal V}\right|_{\rm min} = -1/(4\pi)$,
\begin{equation}
m =  \sqrt{1-2 \mu^2 \nu^2 - 2 \mu \nu \sqrt{\mu^2 \nu^2+ \mu^2 -1}},
\label{B5}
\end{equation}
into Eqn.~(\ref{B3}). For given $\nu$, this enables us to find $\mu_{\rm crit}$ (where the 1st order phase transition happens) numerically.
Examples will be shown below. Let us first check whether chiral symmetry gets restored at any chemical potential. Evaluating the asymptotic form of the effective potential
for large $\mu$ and minimizing it with respect to $m$, we find two solutions
\begin{equation}
m_1=0, \quad m_2= (2 \mu)^{\nu/(\nu-1)} \quad (\mu \gg 1).
\label{B6}
\end{equation}
A comparison of the effective potentials at these two masses gives
\begin{equation}
{\cal V}_{\rm eff}(m_2) -{\cal V}_{\rm eff}(m_1) = \frac{(-1+\nu)}{\pi} 2^{2/(\nu-1)} \mu^{2\nu/(\nu-1)} \quad (\mu \gg 1).
\label{B7}
\end{equation}
This is always negative for $0 < \nu < 1$. Thus chiral symmetry does not get restored in the two-component system. 

For the comparison with the calculations in the main text, we still need the fermion density and energy density. In the region where $\mu>m$,
the fermion density is 
\begin{equation}
\rho = - \frac{\partial}{\partial \mu} {\cal V}_{\rm eff} = \frac{\nu \sqrt{\mu^2-m^2}}{\pi}.
\label{B8}
\end{equation}
The energy density can be obtained from the effective potential as
\begin{eqnarray}
{\cal E} & = &  {\cal V}_{\rm eff} + \mu \rho
\nonumber \\
& = & \frac{m^2}{4\pi}( \ln m^2 - 1) + \frac{\nu}{2\pi} \left\{ m^2 \ln \left(\frac{\mu+\sqrt{\mu^2-m^2}}{m}\right) + \mu \sqrt{\mu^2-m^2} \right\}
\label{B9}
\end{eqnarray}
These are all the equations needed. For given $\mu,\nu$, find $m$ by minimization first, then determine $\rho$ and ${\cal E}$.
Above the phase transition, we then just plot ${\cal E}$ vs. $\rho$. Below the phase transition, there is a mixed phase. This simply means that the critical
point in the (${\cal E},\rho$) where the phase transition happens has to be connected to the vacuum point by a straight line (Maxwell construction). 

\begin{figure}
\begin{center}
\epsfig{file=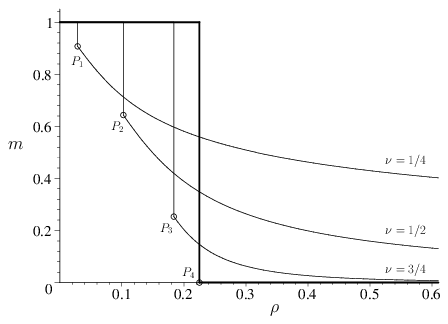,height=6cm,angle=0}
\caption{Order parameter $m$ vs. fermion density for homogeneous two-component calculation and three values of $\nu$. The dynamical mass always starts
at the vacuum value 1 until a first order transition to a smaller value $m_{\rm crit}$, then goes down monotonically without reaching 0. The coordinates of the 
critical points $P_i$ are given in table~\ref{tab1}. The fat, rectangular curve is the result for $\nu=1$, i.e., the standard one-component case with a critical density
corresponding to the critical chemical potential $1/\sqrt{2}$ \cite{L19}.}
\label{fig10}
\end{center}
\end{figure}

We now turn to the results. In Fig.~\ref{fig10}, the order parameter $m$ is plotted against fermion density $\rho$, for a few values of $\nu$.
The fat solid curve actually corresponds to $\nu=1$ where we must recover the result for the standard GN model with one fermion species only.
The phase transition with restoration of chiral symmetry happens at $\mu=1/\sqrt{2}$. With decreasing $\nu$ values the critical density moves
downward and the jump in fermion mass is reduced, see also table~\ref{tab1} for the critical parameters.

\begin{figure}
\begin{center}
\epsfig{file=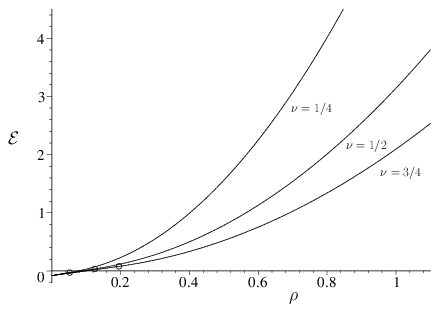,height=6cm,angle=0}
\caption{Result of two-component HF calculation for a homogeneous mean field and three values of $\nu$. The circles correspond to the 1st order
phase transition. On this scale, it is hard to see a difference between this calculation and the one based on the twisted kink crystal, Fig.~\ref{fig3}.
See however the low density close-ups in Figs.~\ref{fig4} and \ref{fig5}.}
\label{fig11}
\end{center}
\end{figure}

Fig.~\ref{fig11} shows the energy density vs. fermion density for the three values of $\nu$ used in the present paper. The circles denote the critical 
points where the mass has a discontinuity. At first glance, these figures are indistinguishable from the full calculation based on the gapped
twisted kink crystal potential. To highlight the differences, close-ups of the region around $\rho=0$ have been included in Figs.~\ref{fig4} and \ref{fig5},
see Sect.~\ref{sec4}.

\begin{center}
\begin{table}
\begin{tabular}{|c|c|c|c|c|}
\hline
 & $\nu$  & $\rho_{\rm crit}$ & $m_{\rm crit}$  & $\mu_{\rm crit}$ \\
\hline
$P_1$ & 1/4 & 0.0288 & 0.9074 & 0.9768 \\
$P_2$ & 1/2 & 0.1024 & 0.6436 & 0.9102 \\
$P_3$ & 3/4 & 0.1838 & 0.2534 & 0.8104 \\
$P_4$  & 1 & 0.2251 & 0 & 0.7071 \\
\hline
\end{tabular}
\caption{Critical data for the points $P_i$ shown in Fig.~\ref{fig10}.}
\label{tab1}
\end{table}
\end{center}

\end{document}